\documentclass[12pt,preprint]{aastex}

\newcommand{\phunit}{photons cm$^{-2}$ sr$^{-1}$ s$^{-1}$ \AA$^{-1}$}

\newcommand{\lya}{Ly~$\alpha$}
\newcommand{\lyb}{Ly~$\beta$}

\newcommand{\fuse}{{\it FUSE}}
\newcommand{\thc}{$\theta^{1}$ Ori C}

\shorttitle{Intense Diffuse Far-UV Emission from the Orion Nebula}
\shortauthors{Murthy et al.}

\begin{document}

\title{Intense Diffuse Far-UV Emission from the Orion Nebula}
\author{Jayant Murthy}
\affil{Indian Institute of Astrophysics, Koramangala, Bangalore 560 034}
\email{jmurthy@yahoo.com}

\author{David J. Sahnow and R. C. Henry}
\affil{The Johns Hopkins University, Baltimore, MD 21218}
\email{sahnow@pha.jhu.edu}
\email{henry@jhu.edu}

\begin{abstract}

We present spectra of the diffuse FUV (900 -- 1200 \AA) emission from a region near the Orion
Nebula: the first high resolution spectra of the diffuse background radiation. These observations
were made using serendipitous \fuse\ observations and were only possible because of the strength of 
the diffuse emission ($\approx 3 \times\ 10^{5}$ \phunit) and the sensitivity of the \fuse\ instrument.
Preliminary modeling suggests that the light is
scattered starlight from the Trapezium stars, primarily \thc. However, a comparison of the spectra 
with nearby stars shows significant differences in the presence and strength of the absorption lines,
particularly in the \lyb\ line where there is
much less interstellar absorption in our diffuse spectrum. We believe that we are indeed observing scattered
light from the Trapezium stars but through a line of sight with much less matter than the direct line
to the stars.

\end{abstract}


\section{Introduction}

The Orion Nebula (M42) is one of the best studied of all astronomical objects with emission in
different wavelengths reflecting different aspects of its physics and morphology. An excellent 
description of the observations and the resulting model of the region is given by \citet{odell01}. 
Intense ultraviolet emission from the Orion Nebula has been observed by rocket flights
(\citet{boh82}, \citet{ona84}) and from the International Ultraviolet Explorer \citep{mat81} and is 
presumed to be due to the forward scattering of the light of the brightest of the 
Trapezium stars --- \thc\ (HD 37022) --- by dust close to and in front of the star. Note that the 
scattering is not actually from the Orion Nebula, which is defined as a region of 
ionized gas in front of the Orion Molecular Cloud (OMC-1), but is rather from dust in the same 
direction. In this
paper we present the first observations of diffuse emission from Orion in the far ultraviolet (FUV)
using serendipitous observations made by the Far Ultraviolet Spectroscopic Explorer (\fuse).

\section{Observations and Data Analysis}

\fuse\ was 
launched on June 24, 1999 into a low Earth orbit (LEO) by a Delta II rocket and has been observing
astronomical targets since then. The spacecraft and mission have been described by \citet{M00} 
and \citet{S00}. The instrument
consists of four coaligned optical channels, two of which are coated with silicon carbide (SiC)
and two with lithium fluoride (LiF) over aluminum, providing coverage over the spectral range from 905 -- 
1187 \AA. Diffuse sources, such as we report on here, will be visible in all 3 apertures (LWRS,
MDRS, and HIRS) but useful spectra can only be obtained from the LWRS ($30\arcsec\ \times\ 30\arcsec$)
and, for sufficiently bright sources, MDRS ($4\arcsec\ \times\ 20\arcsec$) apertures. 

\begin{figure}
\plotone{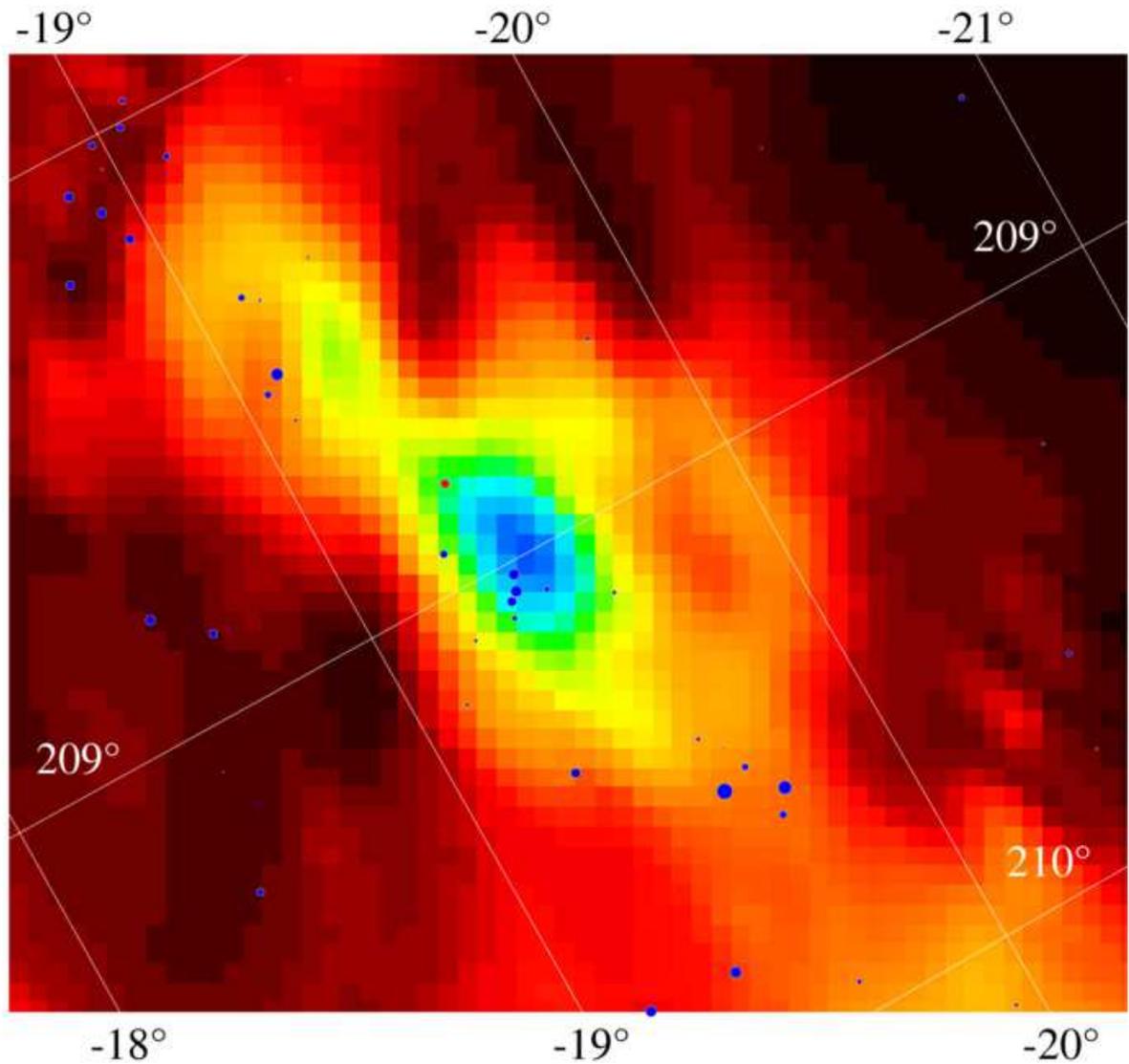}
\caption{An interstellar dust map \citep{S98} is shown with the Orion Nebula (M42) in the center of the image.
Overplotted are the hot stars in Orion with the size of the symbol proportional to the TD-1 magnitude
of the star. The two S40546 observations are very near the star HD36981 (red circle 
at the NE edge of M42). The LWRS 
($30\arcsec \times 30\arcsec$) aperture is centered 105\arcsec\ to the NE of the star and the MDRS
aperture 100\arcsec\ to the SW. The observations are about 12\arcmin\ away from \thc\
which provides most of the power for the surrounding Orion Nebula.}
\label{dss_fig}
\end{figure}

Two of the observations in the S405/505 program, an operational program used as part of the
channel realignment of the \fuse\ spectrographs, were of a region of empty space near the star
HD 36981  (Fig. \ref{dss_fig}, Table \ref{obs_log}). A check of the attitude has shown that the actual pointing
was within a few arcseconds of the nominal position. 
We have downloaded the data from these two observations
and processed them using the standard CALFUSE pipeline (v2.4, Dixon, Kruk, \& Murphy 2002) with 
the modifications described by \citet{mur04}. One of these modifications was to use only the
data from the ``NIGHT'' part of the orbit in order to exclude all airglow emission, apart from
the Lyman lines of atmospheric hydrogen. Although there
may still be residual amounts of the \ion{O}{1} lines around 1040 \AA\ and the \ion{N}{1} lines at 
1134 \AA\ \citep{F01}, these will not contribute to the intense continuum emission observed in Orion.
Other modifications to the standard pipeline were that we added together all the different exposures
which form a single observation and replaced the background subtraction procedure
with one where we estimated the background from the region on either side of the aperture. Finally, in
order to increase the signal-to-noise ratio, we binned together the spectra by a factor of 32 in
wavelength, to reduce the effective resolution to about 700. We have used this procedure to reduce the
data from most of the S405/S505 observations and the results are discussed by \citet{mur04}. This
is the brightest of the diffuse fields observed in that program.

\begin{figure}
\plotone{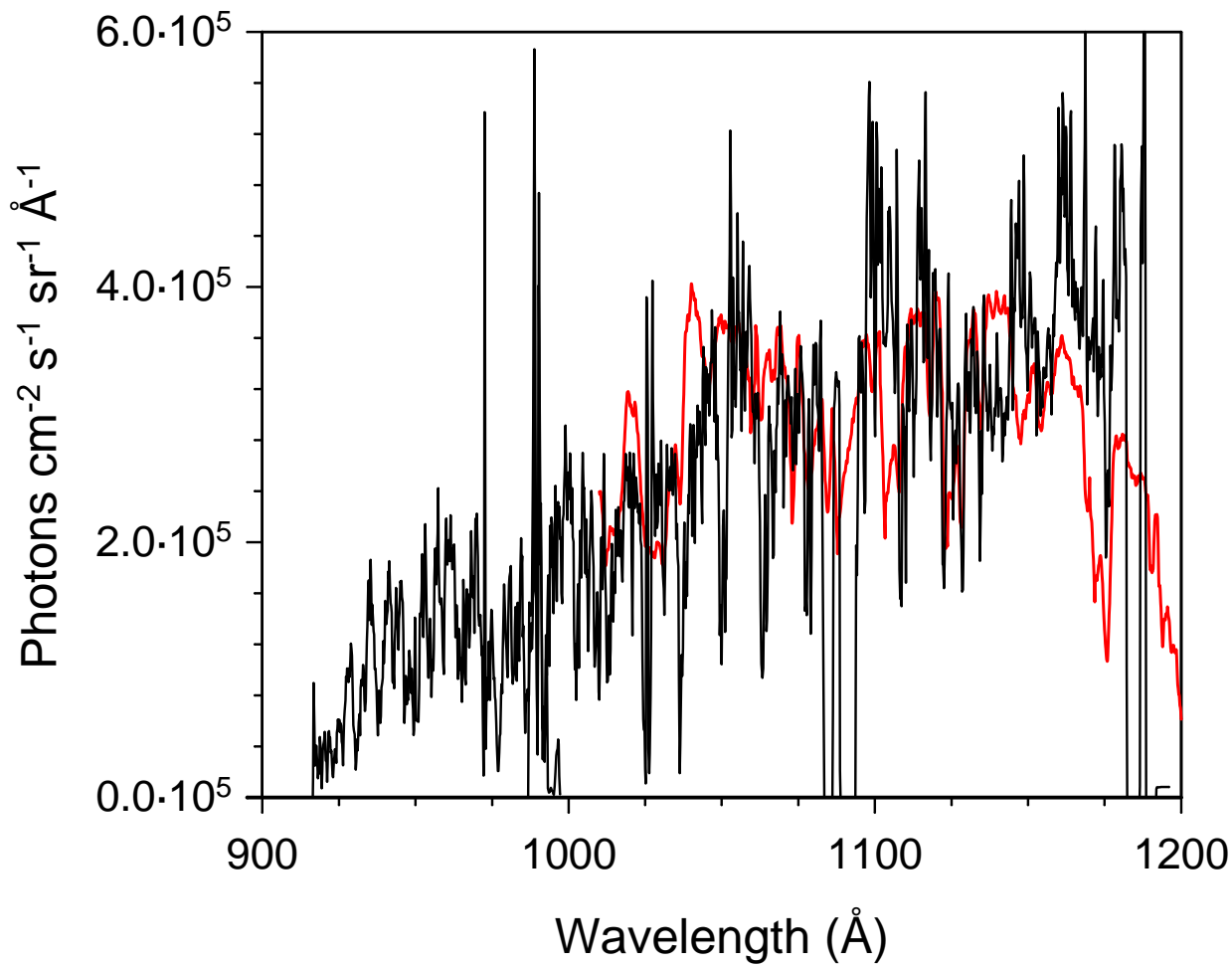}
\caption{The spectrum for our diffuse observation is shown. Only the spectra from three of the 
detector segments are shown for
clarity; the extracted spectra from the other segments are consistent with these. Also omitted for clarity are the error bars which
are on the order of 10000 \phunit. Note the excellent agreement between segments at the edges, validating
our empirical background subtraction procedure. Superimposed on the plot is a scaled
spectrum of $\theta^{1}$ Ori (red line) taken from the Copernicus archive. The spectral shape is
similar to the diffuse spectrum until the onset of interstellar \lya\ absorption near 1200 \AA.
}
\label{spectra}
\end{figure}

The resultant spectra are shown in Fig. \ref{spectra} in which we have coadded the data from the two
different observations. There are three different spectra shown in the Figure, all from the LWRS
aperture: the coadded 1B and 2A SiC spectra (900 -- 1000 \AA); the 1A LiF spectrum (980 -- 1100 \AA);
and the coadded 1B and 2A LiF spectra (1080 -- 1200 \AA). The same data are shown at higher resolution 
in Figs. \ref{lines} and \ref{long_spec}.
For clarity, we have not shown the
spectra from the other \fuse\ segments which are entirely consistent with those plotted here. Because
the diffuse emission is so bright, we were also able to extract spectra from the MDRS aperture which
was pointed about 3\arcmin\ away from the LWRS. Despite the slightly different pointing,
the MDRS spectra were essentially identical with the LWRS spectra.

\section{Results and Discussion}

\begin{figure}
\plotone{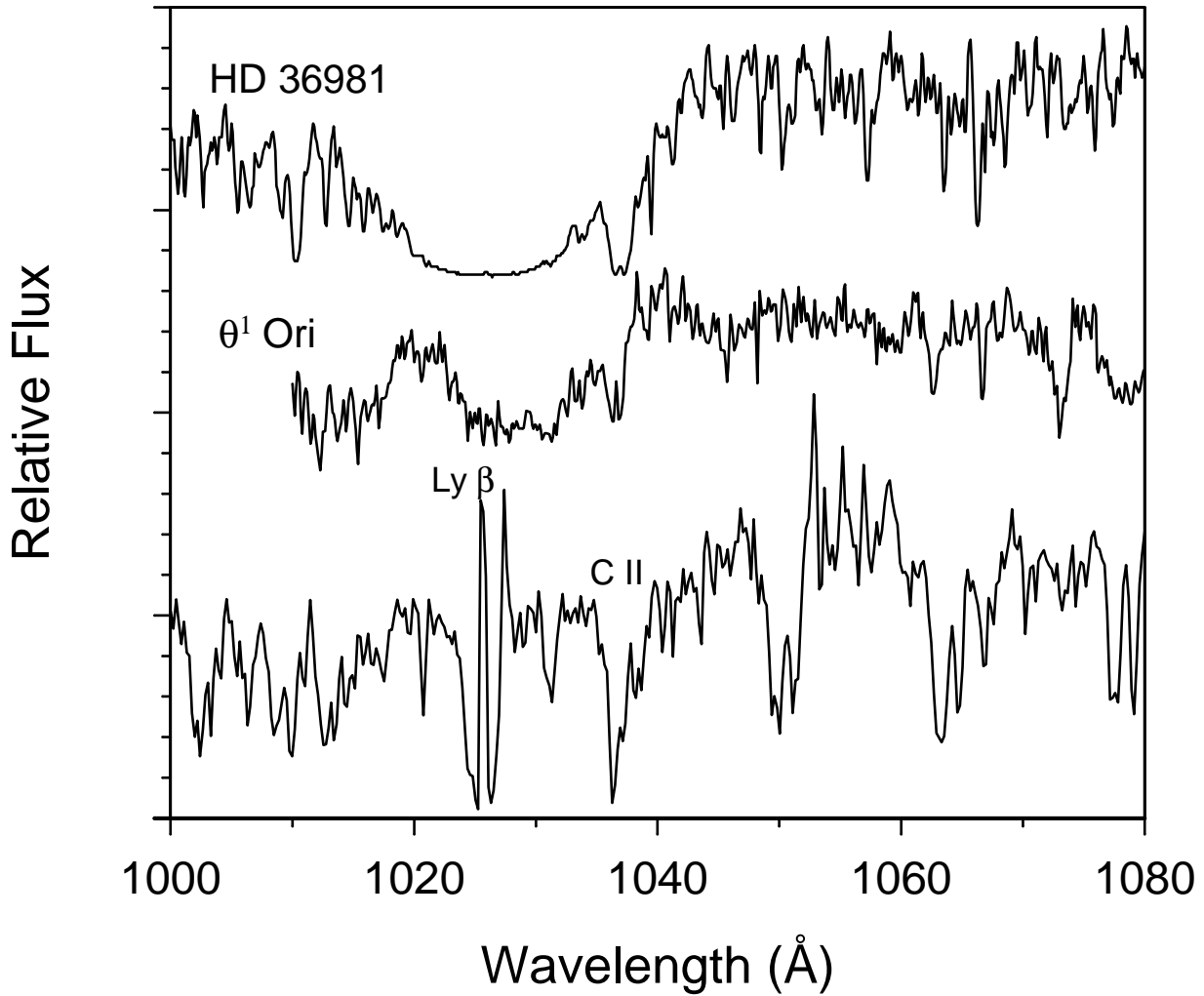}
\caption{
Plotted are (from top to bottom) the \fuse\ spectrum of HD 36981, the Copernicus spectrum of \thc\
and the \fuse\ observation reported here. The \lyb\ line is much broader in both stellar spectra
than in the diffuse observation; in the case of HD 36981 due to self-absorption in the star and
due to interstellar absorption in the case of \thc. The geocoronal \lyb\ emission
line is seen in the middle of the interstellar \lyb\ (1026 \AA) 
absorption line in our diffuse \fuse\ observation. Note that, surprisingly, the diffuse spectrum is
rich in absorption lines which are not seen in either of the stellar spectra. France 
(personal communication) has suggested that these are largely due to H$_2$ absorption but also  with
contributions from flurescent H$_2$ emission; however, a full
analysis is beyond the scope of this work.
}
\label{lines}
\end{figure}

The \fuse\ spectrographs are the first instruments with sufficient sensitivity to 
perform absorption line spectroscopy of the diffuse ISM, albeit only in bright
areas such as Orion. The diffuse emission is due to scattered light from the nearby stars and a
comparison of the spectra can provide important clues about the source of the photons and the 
geometry of the dust. Although our first assumption was that we were observing scattering of
the light of HD 36981 (B5V, V = 8), only about 2\arcmin\ away, a comparison of the spectra near
\lyb\ shows that 
HD 36981 cannot be the major contributor to the scattered radiation because the broad intrinsic 
\lyb\ absorption in the star is not reflected in our observed spectrum (Fig. \ref{lines}).

\begin{figure}
\plotone{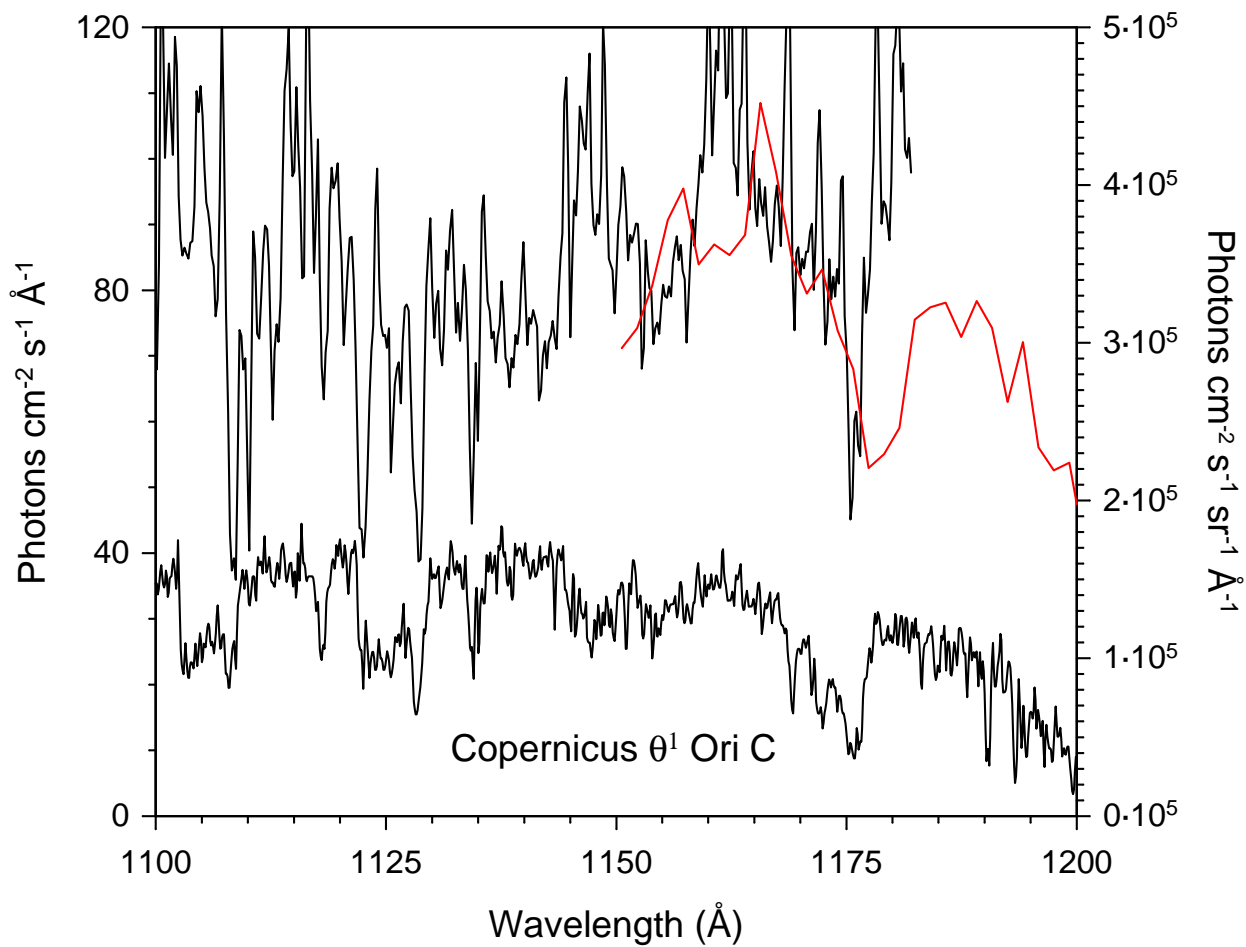}
\caption{
Plotted is our diffuse spectrum at long wavelengths with a calibrated Copernicus spectrum
of \thc\ and the diffuse observation are plotted. Because of an uncertain
calibration of the Copernicus data, we have also plotted an IUE observation of \thc\
which is well calibrated (red line).
The units are, respectively, photons cm$^{-2}$ sr$^{-1}$ s$^{-1}$ for the stellar spectra on the left hand
side and \phunit\ for the diffuse spectrum, on the right hand side.
}
\label{long_spec}
\end{figure}

Despite its much greater distance (12\arcmin) from our observed locations,
\thc\ (HD 37022; O5V; V = 5.1) ---
the brightest of the Trapezium stars --- may contribute as much or more energy as
HD 36981, depending on the relative geometry.  Unfortunately HD 37022 is much too bright to observe with 
\fuse\ but was observed by Copernicus. We downloaded the spectrum from the MAST
archives and applied the calibration of \citet{sno77} and this is plotted in
Fig. \ref{spectra}. The shape matches the diffuse spectrum well until about 1150 \AA\ when interstellar
\lya\ absorption eats into the stellar spectrum. It was difficult to establish
the absolute calibration of the Copernicus spectrograph and instead we 
downloaded an International Ultraviolet Explorer observation (SWP 5085) from MAST to compare
the stellar flux with the diffuse spectrum, at least at wavelengths above 1150 \AA\
(Fig. \ref{long_spec}).

Despite the overall similarity in the shape of the spectra, a detailed comparison
(Figs. \ref{lines} and \ref{long_spec}) shows many differences. As with HD36981,
the \lyb\ absorption line is much broader in the spectrum of \thc\ than in our diffuse spectrum but,
in this case, is due to interstellar absorption in the several interstellar clouds
in front of the Trapezium \citep{pr01}. Many of the other lines in the diffuse spectrum
have been identified as 
being due to molecular hydrogen in both absorption and emission (France - personal communication,
\citet{M03}) which, surprisingly, do not
appear with the same strength in either of the stellar spectra, even though the stars are approximately
co-located with the scattering dust.

While a definitive interpretation must await a more detailed
analysis of the scattering geometry, it seems likely that we are observing the scattered light from
\thc\ through a much lower column density line of sight than the direct line to the star.
We have begun the process of modeling the entire spectrum including the absorption lines in order
to determine the scattering properties of the dust grains near Orion.

\section{Conclusions}

We present the first high resolution spectra of diffuse scattering in the FUV using serendipitous
\fuse\ observations of a blank patch of sky near HD 36981 and just off the Orion Nebula. We 
propose that the emission is due to scattering of the light from \thc\ (12\arcmin\ away) rather than
the much closer HD 36981 (2\arcmin\ away), based on a comparison of the spectra. The HD 36981 spectrum
has broad self-absorbed Lyman absorption lines which are not seen in the diffuse spectrum. Although
there are also broad Lyman absorption lines in the spectrum of \thc, these are due to interstellar
absorption in foreground gas and suggest that the path traversed by the scattered photons includes
much less gas than the direct path to the star.

We believe that such high resolution observations of the diffuse scattering provide an important new tool
for the study of interstellar dust scattering not just in understanding the spectral properties of the
interstellar dust but also in providing clues to the origin of the diffuse emission. We have
proposed new \fuse\ observations of several locations in and around M42 and hope to integrate these
observations with archival {\it IUE} data in the region \citep[for example, ][]{mat81}. Through these
data and making use of the exceptionally well understood geometry around Orion, we should be able
to, for the first time, pin down the optical properties of the interstellar dust grains from the
FUV through the visible.

\acknowledgments
We thank the FUSE team for their support and quick response to the many questions we
raised while working on this problem. In particular, Bill Blair, Van Dixon, Alex Fullerton
and B-G Andersson provided clarification on many points. Steve McCandliss and Kevin France
pointed out the 
importance of H$_2$ absorption and emission in the diffuse spectrum. The data presented in this paper
was obtained from the Multimission Archive at the Space Telescope Science Institute (MAST). 
STScI is operated by the Association of Universities for Research in Astronomy, Inc., 
under NASA contract NAS5-26555. Support for MAST for non-HST data is provided by the 
NASA Office of Space Science via grant NAG5-7584 and by other grants and contracts. This
research has made use of NASA's Astrophysics Data System and the SIMBAD database, operated at
CDS, Strasbourg, France.

\clearpage

\clearpage

\begin{deluxetable}{rcccc}
\tablecaption{Observational Parameters}
\tablewidth{0pt}
\tablehead{
\colhead{Observation } & \colhead{Aperture} & \colhead{Night Exposure Time (s)}   & \colhead{L} & \colhead{B}}
\startdata
S4054601 & LWRS & 10565 & 208.81 & -19.31\\
S4054601 & MDRS & 10565 & 208.81 & -19.31\\
S4054602 & LWRS & 5696 & 208.82 & -19.36\\
S4054602 & MDRS & 5696 & 208.81 & -19.36\\
\enddata
\label{obs_log}
\end{deluxetable}

\end{document}